\documentclass[%
superscriptaddress,
nofootinbib,
nobibnotes,
amsmath,amssymb,
aps,
floatfix,
]{revtex4-2}

\usepackage{placeins}
\usepackage{booktabs}
\usepackage{dcolumn}
\usepackage{array}
\usepackage{longtable}
\usepackage{float}
\usepackage{hyphenat}
\usepackage{natbib}
\usepackage{hyperref}

\usepackage{amsmath,amsfonts,amsthm,amssymb}
\usepackage{bm,graphicx,graphics,color}
\usepackage{epsf,epsfig}
\usepackage[all]{xy}
\newcommand*{\B}[1]{\ifmmode\bm{#1}\else\textbf{#1}\fi}

\def\no{\nonumber}
\def\lb{\label}
\def\be{\begin{equation}}
\def\ee#1{\label{#1}\end{equation}}
\newcommand{\ben}{\begin{eqnarray}}
\newcommand{\een}{\end{eqnarray}}

\begin{document}

\title{Post-Newtonian Spherically Symmetrical Accretion}

\author{Gilberto M. Kremer}
\email{kremer@fisica.ufpr.br}
\affiliation{Departamento de F\'{i}sica, Universidade Federal do Paran\'{a}, Curitiba 81531-980, Brazil}
\author{Leandro C. Mehret}
\email{lcmehret@gmail.com}
\affiliation{Departamento de F\'{i}sica, Universidade Federal do Paran\'{a}, Curitiba 81531-980, Brazil}

\begin{abstract}
The objetive of this work is to investigate the influence of the corrections to the spherical symmetrical accretion of an infinity  gas cloud  characterized by a polytropic equation  into a massive object due to the post-Newtonian approximation. Starting with the steady state post-Newtonian hydrodynamic equations for the mass, mass-energy and momentum densities, the post-Newtonian Bernoulli equation is derived. The post-Newtonian corrections to the critical values of the flow velocity, sound velocity and radial distance are obtained from the system of hydrodynamics equations in spherical coordinates. It was considered that the ratio of the sound velocity far the massive body and the speed of light was of order $a_\infty/c=10^{-2}$. The analysis of the solution led to following results: the Mach number for the Newtonian and post-Newtonian accretion have practically the same values for radial distances of order of the critical radial distance; by decreasing the radial distance the Mach number for the Newtonian accretion is bigger than the one for the  post-Newtonian accretion;  the difference between the Newtonian and post-Newtonian Mach numbers when the ratio $a_\infty/c\ll10^{-2}$ is insignificant; the effect of the correction terms in post-Newtonian Bernoulli equation is more perceptive for the lowest values of the radial distance; the solutions for $a_\infty/c>10^{-2}$ does not lead to a continuous inflow and outflow velocity at the critical point;
the comparison of the solutions with those that follow from a relativistic Bernoulli equation shows that the dependence of the Mach number with the radial distance of the former is bigger than the Newtonian and post-Newtonian Mach numbers.
\end{abstract}
\maketitle

\section{Introduction}

An important area of research in astrophysics is related to the steady state problem of spherically symmetrical accretion  of a perfect gas into a massive object. The pioneers works in this subject were published  by  Hoyle and Lyttleton \cite{1,2}, Bondi \cite{3,4} and Michel \cite{5}. Nowadays this problem is still a subject of several investigations where a relativistic fluid accretes into a massive body described by other metrics: Schwarzschild -- de Sitter, Reissner-Nordstr\"om, Reissner-Nordstr\"om -- de Sitter  (see e.g. \cite{6,7,8,9,10,11,12,13,14,15,16,17,18,19,20} and the references therein).

The aim of this paper is to investigate the influence of the corrections to the Newtonian accretion due to the post-Newtonian approximation. We start from the post-Newtonian hydrodynamic equations for the mass, mass-energy and momentum densities and derive the post-Newtonian Bernoulli equation. The post-Newtonian corrections to the critical values of the flow velocity, sound velocity and radial distance are obtained from the system of hydrodynamics equations. It is shown that due to the post-Newtonian corrections the critical point does not correspond to the transonic point as in the Newtonian accretion. The solution for the Mach number as function of a dimensionless radial distance depends on the ratio of the sound velocity far the massive body and the speed of light $a_\infty/c$. This ratio was fixed to be  $a_\infty/c=10^{-2}$ which is of relativistic order. For values of this ratio greater than $10^{-2}$ there is no continuity in the inflow and outflow velocity at the critical point, while for values smaller than $10^{-2}$ there is no difference between the Newtonian and post-Newtonian solutions. For $a_\infty/c=10^{-2}$   the Mach number for the Newtonian and post-Newtonian accretion have practically the same values for radial distances of order of the critical radial distance, but by decreasing the radial distance the Mach number for the Newtonian accretion is bigger than the one for the  post-Newtonian accretion.    The weak field limit of the  relativistic Bernoulli equation [5] is also developed  and a comparison of the solutions are investigated.

The paper is organized as follows: in Section \ref{sec2} the post-Newtonian hydrodynamic equations are introduced, while in Section \ref{sec3} the post-Newtonian mass density accretion rate, the Bernoulli equation and the critical values for the flow velocity, sound speed and radial distance are obtained. In section \ref{sec4} the post-Newtonian equation for the Mach number as function of a dimensionless radial distance is deduced. In Section \ref{sec5} the relativistic spherically symmetrical accretion  is developed. The analysis of the Newtonian and post-Newtonian solutions is developed in Section \ref{sec6}. We close the paper with a summary of the results in Section \ref{sec7}.

\section{Post-Newtonian Hydrodynamic equations}\lb{sec2}
The post-Newtonian approximation is  a method of successive approximations in $1/c^2$ powers of the light speed for the  solution of Einstein's field equations. It was proposed in 1938 by  Einstein, Infeld and Hoffmann  \cite{Eins} and the corresponding hydrodynamic equations in the first post-Newtonian approximation (1PN) were obtained by Chandrasekhar \cite{Ch1,Ch1a}.

 In the  post-Newtonian approximation Einstein's field equations are solved for an Eulerian fluid characterized by the energy-momentum tensor
\ben\lb{pf}
T^{\mu\nu}=(\epsilon+p)\frac{U^\mu U^\nu}{c^2}-pg^{\mu\nu}.
\een
Here $U^\mu$ denotes the four-velocity (with $U^\mu U_\mu=c^2$), $g^{\mu\nu}$ the metric tensor while $\epsilon$ and $p$ the energy density and pressure of the fluid, respectively. The energy density   has two parts $\epsilon=\rho c^2(1+\varepsilon/c^2)$ one associated with the mass density $\rho=mn$ and another to the internal energy density $\varepsilon$.  The internal energy density for  a non-relativistic perfect fluid is given by $\varepsilon=p/(\gamma-1)\rho$, where $\gamma=c_p/c_v$ is the ratio of the specific heats at constant pressure and constant volume. For a fluid of monatomic molecules $c_v=3k/2m$ with $k$ denoting Boltzmann constant  and $m$ the rest mass of a fluid molecule.

 The solution of Einstein's field equations leads to the following  components of the metric tensor
\ben\lb{1a}
&&g_{00}=1+\frac{2\phi}{c^2}+\frac2{c^4}\left(\phi^2+\psi\right)+\mathcal{O}(c^{-6}),
\\\lb{1b}
&&g_{0i}=-\frac1{c^3}\zeta_i+\mathcal{O}(c^{-5}),
\\\lb{1c}
&&g_{ij}=-\left(1-\frac{2\phi}{c^2}\right)\delta_{ij}+\mathcal{O}(c^{-4}),
\een
while the corresponding components of the four-velocity in 1PN are
\ben\lb{2}
U^0=c\bigg[1+\frac1{c^2}\bigg(\frac{V^2}2-\phi\bigg)+\frac1{c^4}\bigg(\frac{3V^4}8-\frac{5\phi V^2}2+\frac{\phi^2}2-\psi+\zeta_iV_i\bigg)\bigg],
\een
with $ U^i=U^0V_i/c$.
Above $\phi$ is the Newtonian gravitational potential which satisfies the Poisson equation $\nabla^2\phi=4\pi G\rho$. The corresponding Poisson equations for the scalar $\psi$ and the vector $\zeta_i$ gravitational potentials, read
\ben\lb{3a}
&&\nabla^2\zeta_i=16\pi G\rho V_i,
\\\lb{3b}
 &&\nabla^2\psi=8\pi G\rho\left(V^2-\phi+\frac\varepsilon2+\frac{3p}{2\rho}\right)+\frac{\partial^2 \phi}{\partial t^2},
\een
The gravitational potentials $\phi$, $\zeta_i$ and $\psi$ are those introduced by Weinberg \cite{Wein} and their  connection with the gravitational potentials $U$, $U_i$ and $\Phi$ of Chandrasekhar \cite{Ch1} are
\ben\lb{4}
 \phi=-U,\quad\zeta_i=-4U_i+\frac12\frac{\partial^2\chi}{\partial t\partial x^i},\quad \psi=-2\Phi,
\een
where $\chi$ is a super-potential which obeys the equation $\nabla^2\chi=-2U$.

The hydrodynamic equation for the mass density is obtained from the particle four-flow balance law ${N^\mu}_{;\mu}=0$ together with the representation $N^\mu=nU^\mu$ where $n$ denotes the particle number density, yielding
\ben\lb{5}
\frac{\partial\rho_*}{\partial t}+\frac{\partial \rho_*V_i}{\partial x^i}=0,\qquad\hbox{where}
\qquad
\rho_*=\rho\left[1+\frac1{c^2}\left(\frac{V^2}2-3\phi\right)\right].
 \een
 This equation is the 1PN approximation of the continuity equation and corresponds to eq. \emph{(117)} of Chandrasekhar \cite{Ch1}.

 The mass-energy density hydrodynamic equation in the 1PN approximation follows from the time component of the energy-momentum tensor balance law ${T^{0\nu}}_{;\nu}=0$, resulting
 \begin{equation}\lb{6}
\frac{\partial\sigma}{\partial t}
+\frac{\partial\sigma V_i}{\partial x^i}
-\frac1{c^2}\left(\rho\frac{\partial\phi}{\partial t}+\frac{\partial p}{\partial t}\right)=0,\quad\hbox{where}
\quad
\sigma =\rho\bigg[1+\frac1{c^2}\bigg(V^2-2\phi+\varepsilon+\frac{p}{\rho}\bigg)\bigg].
\end{equation}
The above equation corresponds to eq. \emph{(9.8.14)} of Weinberg \cite{Wein} and eq. \emph{(64)} of Chandrasekhar \cite{Ch1}. Note that we have to identify $\epsilon$ with $\rho$  in Weinberg's book  \cite{Wein} and take $c=1$.

From the spatial components of the energy-momentum tensor balance law ${T^{i\nu}}_{;\nu}=0$ follows the 1PN approximation for the momentum density  hydrodynamic equation, namely
 \ben\no
\frac{\partial\sigma V_i}{\partial t}+\frac{\partial\sigma V_i V_j}{\partial x^j}+\frac{\partial}{\partial x^i}\left[p\left(1-\frac{2\phi}{c^2}\right)\right]+\rho\frac{\partial\phi}{\partial x^i}\left[1+\frac2{c^2}\left(V^2-\phi+\frac\varepsilon2+\frac32\frac{p}\rho\right)\right]
\\\lb{7}
-\frac{4\rho}{c^2}\left(\frac{\partial\phi V_i}{\partial t}+V_j\frac{\partial\phi V_i}{\partial x^j}\right)+\frac\rho{c^2}\frac{\partial\psi}{\partial x^i}
+\frac\rho{c^2}\left(\frac{\partial \zeta_i}{\partial t}+V_j\frac{\partial \zeta_i}{\partial x^j}-V_j\frac{\partial \zeta_j}{\partial x^i}\right)
=0,\quad
\een
which matches eq. \emph{(9.8.15)} of Weinberg \cite{Wein} and eq. \emph{(68)} of Chandrasekhar \cite{Ch1}.

In the Newtonian limiting case the hydrodynamic equations for the mass density (\ref{5}) and the mass-energy density (\ref{6}) coincide and become the continuity equation
\ben\lb{8a}
\frac{\partial\rho}{\partial t}+\frac{\partial \rho V_i}{\partial x^i}=0, \een
while  eq. (\ref{7}) reduces to the Eulerian momentum density hydrodynamic equation
 \ben\lb{8b}
\frac{\partial\rho V_i}{\partial t}+\frac{\partial\rho V_i V_j}{\partial x^j}+\frac{\partial p}{\partial x^i}+\rho\frac{\partial\phi}{\partial x^i}=0.
\een

\section{Post-Newtonian Accretion}\lb{sec3}
\subsection{Post-Newtonian Bernoulli Equation}

In the analysis of the spherically symmetrical accretion a massive object of mass $M$ is surrounded by an infinite gas cloud and is moving with a velocity $V$ relative to it. The gas cloud at large distances from the star is at rest with uniform density and pressure denoted by $\rho_\infty$ and $p_\infty$, respectively. The gas motion is steady spherically symmetrical and it is not taken into account the increase in the  massive object. The gas is characterized by a polytropic equation of state and by a sound velocity $a$ given by
\ben\lb{s1}
p=\kappa \rho^\gamma,\qquad a=\sqrt{\frac {dp} {d\rho}}=\sqrt{\frac{\gamma p}\rho},
\een
where $\kappa$ is a constant and $\gamma$ is related to the polytropic index $n$ by $\gamma=(n+1)/n$.

For steady  states the hydrodynamic equations for mass density (\ref{5}), mass-energy density (\ref{6}) and momentum density (\ref{7}) become
\ben\lb{s2a}
&&\frac{\partial \rho_*V_i}{\partial x^i}=0,
\\\lb{s2b}
&& \frac{\partial\sigma V_i}{\partial x^i}=0,
\\\no
&&\sigma V_j\frac{\partial V_i}{\partial x^j}+\rho\frac{\partial \phi}{\partial x^i}\left[1+\frac2{c^2}\left(V^2-\phi+\frac{3\gamma-2} {2(\gamma-1)}\frac{p}\rho\right)\right]+\frac{\partial}{\partial x^i}\left[p\left(1-\frac{2\phi}{c^2}\right)\right]
\\\lb{s2c}
&&\qquad-\frac{4\rho}{c^2}V_j\frac{\partial \phi V_i}{\partial x^j}+\frac\rho{c^2}\frac{\partial \psi}{\partial x^i}+\frac\rho{c^2}V_j\left(\frac{\partial \zeta_i}{\partial x^j}-\frac{\partial \zeta_j}{\partial x^i}\right)=0.
\een
In the steady state momentum density hydrodynamic equation  (\ref{s2c})  we have used  the corresponding  mass-energy density hydrodynamic equation (\ref{s2b}) and the relationship $\varepsilon=p/(\gamma-1)\rho$.

In spherical coordinates  the fields depend only on the radial coordinate $r$  and  due to the fact that we are dealing with a spherically symmetrical flow, the components of the hydrodynamic velocity are $V_i=(V(r),0,0)$. Hence,  equations  and (\ref{s2a}) -- (\ref{s2c}) become
\ben\lb{s3a}
&&\frac{d\left\{r^2\rho\left[1+\frac1{c^2}\left(\frac{V^2}2-3\phi\right)\right]V\right\}}{dr}=0,
\\\lb{s3b}
&&\frac{d\left\{r^2\rho\left[1+\frac1{c^2}\left(V^2-2\phi+\frac\gamma{\gamma-1}\frac{p}{\rho}\right)\right] V\right\}}{dr}=0,
\\\no
&&\rho\left[1+\frac1{c^2}\left(V^2-6\phi+\frac\gamma{\gamma-1}\frac{p}{\rho}\right)\right] V\frac{d V}{dr}+\frac{d}{dr}\left[p\left(1-\frac{2\phi}{c^2}\right)\right]
\\\lb{s3c}
&&\qquad+\rho\frac{d \phi}{dr}\left[1-\frac2{c^2}\left(V^2+\phi-\frac{3\gamma-2}{2(\gamma-1)}\frac{p}{\rho}\right)\right]+\frac\rho{c^2}\frac{d \psi}{dr}=0.
\een

For the analysis of the flow velocity it is more convenient to introduce the proper velocity of the flow $v_r$ which is measured by a local stationary observer (see e.g \cite{ST,BFMM}). The proper velocity  is defined by
\ben\lb{nn1}
v_r=\frac{U^r}{U_0/c}=\frac{U^r}{(U^0/c)\left(1+2\phi/c^2\right)}.
\een
Since $V_i=U^i/(U^0/c)$, we have that
\ben\lb{vr}
v_r=\frac{V}{\left(1+2\phi/c^2\right)}, \qquad\Longrightarrow\qquad V=v_r\left(1+2\phi/c^2\right).
\een

By taking into account the relationship (\ref{vr})  the system of differential equations (\ref{s3a}) -- (\ref{s3c}) can be rewritten   in terms of the proper velocity $v_r$  as
\ben\lb{nn4a}
&&\frac{d\left\{r^2\rho\left[1+\frac1{c^2}\left(\frac{v_r^2}2-\phi\right)\right]v_r\right\}}{dr}=0,
\\\lb{nn4b}
&&\frac{d\left\{r^2\rho\left[1+\frac1{c^2}\left(v_r^2+\frac{a^2}{\gamma-1}\right)\right] v_r\right\}}{dr}=0,
\\\no
&&\left[1+\frac1{c^2}\left(v_r^2-2\phi+\frac{a^2}{\gamma-1}\right)\right] v_r\frac{d v_r}{dr}+\frac{d\rho}{dr}\frac{a^2}\rho\left(1-\frac{2\phi}{c^2}\right)
\\\lb{nn4c}
&&\qquad+\frac{d \phi}{dr}\left[1-\frac1{c^2}\left(2\phi-\frac{a^2}{\gamma-1}\right)\right]+\frac1{c^2}\frac{d \psi}{dr}=0.
\een
Above we have used to the expression for the sound speed (\ref{s1}).

The integration of (\ref{nn4a}) and (\ref{nn4b}) imply the mass density and mass-energy accretion rates
\ben\lb{s4a}
&&\dot M_{\rho_*}=4\pi \rho r^2\left[1+\frac1{c^2}\left(\frac{v_r^2}2-\phi\right)\right]v_r,
\\\lb{s4b}
&&\dot M_\sigma=4\pi \rho r^2\left[1+\frac1{c^2}\left(v_r^2+\frac{a^2}{\gamma-1}\right)\right] v_r.
\een
From these equations we obtain a relationship between both accretion rates
\ben\lb{s5a}
\dot M_\sigma=\dot M_{\rho_*}\left[1+\frac1{c^2}\underline{\left(\frac{v_r^2}2+\phi+\frac{a^2}{\gamma-1}\right)}\right].
\een
 Here we can use the Newtonian Bernoulli equation
\ben\lb{s5b}
\frac{v_r^2}2+\frac{a^2}{\gamma-1}+\phi=\frac{a_\infty^2}{\gamma-1},
\een
for the underlined term,  since it is of order $1/c^2$ in  (\ref{s5a}, resulting
\ben\lb{s5c}
\dot M_\sigma=\dot M_{\rho_*}\left[1+\frac1{\gamma-1}\frac{a_\infty^2}{c^2}\right].
\een
Hence, the mass density and mass-energy accretion rates differ from each other by a $1/c^2$ term, i.e., in the Newtonian limiting case both coincide, i.e., $\dot M_\sigma=\dot M_{\rho_*}=4\pi \rho r^2V$.

The multiplication of the momentum density (\ref{nn4c}) by
$$\frac1\rho\left[1+\frac1{c^2}\left(2\phi-\frac{a^2}{\gamma-1}\right)\right]$$ leads to the following differential equation
\ben\lb{s6}
 v_r\frac{dv_r}{dr}\left[1+\frac{v_r^2}{c^2}\right]+\frac{d\rho}{dr}\frac{a^2}\rho\left[1-\frac{a^2}{c^2(\gamma-1)}\right]
+\frac{d\phi}{dr}+\frac1{c^2}\frac{d\psi}{dr}=0.
\een

 The post-Newtonian Bernoulli equation follows from the integration of (\ref{s6}), yielding
\ben\lb{s7}
\frac{v_r^2}2\bigg[1+\frac{v_r^2}{2c^2}\bigg]
+\frac{a^2}{\gamma-1}\left[1-\frac{a^2}{2c^2(\gamma-1)}\right]
+\phi+\frac{\psi}{c^2}=\frac{a_\infty^2}{\gamma-1}\left[1-\frac{a_\infty^2}{2c^2(\gamma-1)}\right].
\een
Here we have assumed that the proper velocity  $v_r$ and the gravitational potentials $\phi,\psi$ vanish far from the massive object. Note that (\ref{s7}) reduces to the Newtonian Bernoulli equation (\ref{s5b}) by neglecting the terms in $1/c^2$.

We can rewrite the hydrodynamic equations in spherical coordinates (\ref{nn4a}), (\ref{nn4b}) and (\ref{s6}) as
\ben\lb{s8a}
&&\frac2r+\frac{\rho'}\rho+\frac{v_r'}{v_r}+\frac1{c^2}\left(v_rv_r'
-\phi'\right)=0,
\\\lb{s8b}
&&\frac2r+\frac{\rho'}\rho+\frac{v_r'}{v_r}+\frac1{c^2}\left(2v_rv'_r+a^2\frac{\rho'}\rho\right)=0,
\\\lb{s8c}
&&\left(1+\frac{v_r^2}{c^2}\right)v_rv_r'+\left[1-\frac{a^2}{c^2(\gamma-1)}\right]a^2\frac{\rho'}\rho
+\phi'+\frac{\psi'}{c^2}=0,
\een
where the prime denotes the differentiation with respect to $r$.

The system of differential equations  (\ref{s8a}) -- (\ref{s8c}) can be solved as a  system of algebraic equations for $V'$, $\rho'$ and $\psi'$, yielding
\ben\lb{s9a}
&&\frac{v_r'}{v_r}=\frac{2}{r}\frac{a^2\left(1-\frac{r\phi'}{2c^2}\right)-\frac{r\phi'}{2}}{v_r^2\left(1-\frac{a^2}{c^2}\right)-a^2},
\qquad\frac{\rho'}\rho=-\frac2{r}\frac{v_r^2\left(1-\frac{r\phi'}{c^2}\right)-\frac{r\phi'}2}{v_r^2\left(1-\frac{a^2}{c^2}\right)-a^2},
\\\lb{s9c}
&&\psi'=\frac{a^4\left[r\phi'-2v_r^2\left(1-\frac{r\phi'}{c^2}\right)\right]-2a^2(\gamma-1)v_r^4\left(1-\frac{r\phi'}{2c^2}\right)+r\phi'v_r^4(\gamma-1)}{r(\gamma-1)\left[v_r^2\left(1-\frac{a^2}{c^2}\right)-a^2\right]}.
\een

We infer from  (\ref{s9a})  that the solution must pass through a   critical point defined by a critical radius $r_c$, a critical proper velocity $V_c$ and a critical sound velocity $a_c$ when the nominator and denominator of these equations vanish. The existence of a critical point prevent singularities in the flow solution and guarantees  a smooth monotonic increase of the flow velocity when $r$ decreases. At the critical point we have
\ben\lb{s10a}
&&a_c^2=\frac{r_c\phi'_c}{2\left(1-\frac{r_c\phi'_c}{2c^2}\right)}\approx\frac{r_c\phi'_c}2\left(1+\frac{r_c\phi'_c}{2c^2}\right),
\\\lb{s10b}
 &&V_c^2=\frac{a_c^2}{\left(1-\frac{a_c^2}{c^2}\right)}\approx a_c^2\left(1+\frac{a_c^2}{c^2}\right),
\\\lb{s10c}
&&V_c^2=\frac{r_c\phi'_c}{2\left(1-\frac{r_c\phi'_c}{c^2}\right)}\approx\frac{r_c\phi'_c}2\left(1+\frac{r_c\phi'_c}{c^2}\right).
\een
The above approximations  are valid since we are working with a first post-Newtonian theory.

The value of $\psi'_c$ at the critical point is obtained from the substitution of (\ref{s10c}) into (\ref{s9c}) yielding
\ben\lb{s11a}
\frac{\psi'_c}{c^2}=\frac{\phi_c^2}{2c^2r_c}=\frac{G^2M^2}{2c^2r_c^3},
\een
by taking  into account the expression for the Newtonian gravitational potential
\ben\lb{s11b}
\phi=-\frac{GM}{r}=-r\phi'.
\een

Another way to determine $\psi'$ is to observe that this potential is of order $1/c^2$ and we can approximate (\ref{s9c}) by
\ben\lb{s12a}
\frac{\psi'}{c^2}=\frac{-a^4\left[\phi+2v_r^2\right]-2a^2(\gamma-1)v_r^4-\phi v_r
^4(\gamma-1)}{c^2r(\gamma-1)\left[v_r^2-a^2\right]},
\een
by neglecting the terms proportional to $1/c^2$ and considering the relationship for the Newtonian potential $\phi'r=-\phi$. Now by taking into account the virial theorem $2K+W=0$, where $K$ and $W$ represent the kinetic and potential energies, we can assume that $2v_r^2+\phi=0$ and (\ref{s12a}) reduces to
\ben\lb{s12b}
\frac{\psi'}{c^2}=\frac{\phi^2}{2c^2r}.
\een
The gravitational potential $\psi$ is obtained from the integration of (\ref{s12b}) by using the Newtonian gravitational potential (\ref{s11b}) resulting
\ben\lb{s12c}
\frac{\psi}{c^2}=-\frac{G^2M^2}{4c^2 r^2}=-\frac{\phi^2}{4c^2},\qquad\hbox{so that}
\qquad
\frac{\psi_c}{c^2}=-\frac{\phi_c^2}{4c^2}.
\een
Above it was considered that $\psi$ vanishes at $r\rightarrow\infty$. Note that  (\ref{s12c})$_2$ is the integral of (\ref{s11a}) with respect to $r_c$.

For the determination of the critical values we make use of Bernoulli equation  (\ref{s7}) and the expressions of  the sound speed  (\ref{s10a}),  proper velocity  (\ref{s10c}) and  gravitational potential  (\ref{s12c})$_2$. The resulting equation is  an algebraic equation for the determination of $\phi_c$ at  the critical point. Its value up to order $1/c^{2}$ is
\ben\lb{s13a}
\phi_c=-\frac{4a_\infty^2}{(5-3\gamma)}\left[1+\frac{15-11\gamma}{8(\gamma-1)(5-3\gamma)}\frac{a_\infty^2}{c^2}\right],
\een
which  implies the expression for the critical radius
\ben\lb{s13b}
r_c=\frac{(5-3\gamma)}4\frac{GM}{a_\infty^2}\left[1-\frac{15-11\gamma}{8(\gamma-1)(5-3\gamma)}\frac{a_\infty^2}{c^2}\right],
\een
thanks to the relationship $\phi_c=-GM/r_c$.

The critical values of the sound speed $a_c$ and proper velocity $V_c$ are obtained from  (\ref{s10a}) and  (\ref{s10c}) by using (\ref{s13a}) for the elimination of $\phi'_c=-\phi_c/r_c$, resulting
\ben\lb{s14a}
&&a_c^2=\frac{2a_\infty^2}{(5-3\gamma)}\left[1-\frac{1-5\gamma}{8(\gamma-1)(5-3\gamma)}\frac{a_\infty^2}{c^2}\right],
\\\lb{s14b}
&&V_c^2=\frac{2a_\infty^2}{(5-3\gamma)}\left[1-\frac{17-21\gamma}{8(\gamma-1)(5-3\gamma)}\frac{a_\infty^2}{c^2}\right].
\een

Furthermore, by using the expression for the sound speed $a^2=\gamma p/\rho$, the polytropic equation of state $p=\kappa\rho^\gamma$  and  the critical value  for the  sound speed  (\ref{s14a}) it follows the  critical value of the mass density
\ben\lb{s15a}
\frac{\rho_c}{\rho_\infty}=\left(\frac{a_c}{a_\infty}\right)^\frac2{\gamma-1}=\left(\frac2{5-3\gamma}\right)^\frac1{\gamma-1}
\left[1-\frac{1-5\gamma}{8(\gamma-1)^2(5-3\gamma)}\frac{a_\infty^2}{c^2}\right].
\een

The mass-density accretion rate (\ref{s4a}) can be written  as
\ben\lb{s15b1}
\dot M_{\rho_*}=4\pi \rho_c r_c^2\left[1+\frac1{c^2}\left(\frac{V_c^2}2-\phi_c\right)\right]V_c
=4\pi \lambda_c\left(\frac{GM}{a_\infty^2}\right)^2\rho_\infty a_\infty,
\een
where the so-called  critical accretion eigenvalue \cite{ST}  is given by
\ben\lb{s15b}
\lambda_c=\left(5-3\gamma\right)^\frac{3\gamma-5}{2\gamma-2}2^\frac{9-7\gamma}{2\gamma-2}\left[1+\frac{121-216\gamma+103\gamma^2}{16(5-3\gamma)(\gamma-1)^2}
\frac{a_\infty^2}{c^2}\right].
\een

In the Newtonian limiting case (\ref{s13a}) -- (\ref{s15b}) reduce to the well-know values found in the literature (see e.g. \cite{ST}), namely
\ben\lb{s16}
&&\phi_c=-\frac{4a_\infty^2}{(5-3\gamma)},\qquad r_c=\frac{(5-3\gamma)}4\frac{GM}{a_\infty^2},
\\
&&a_c^2=V_c^2=\frac{2a_\infty^2}{(5-3\gamma)}, \qquad \frac{\rho_c}{\rho_\infty}=\left(\frac2{5-3\gamma}\right)^\frac1{\gamma-1},
\\
&&\lambda_c=\left(5-3\gamma\right)^\frac{3\gamma-5}{2\gamma-2}2^\frac{9-7\gamma}{2\gamma-2}.
\een

\subsection{Mach number as Function of the Radial Distance}\lb{sec4}

For the determination of the dependence of the flow velocity as function of the radial distance  Bondi \cite{4}  introduced the following dimensionless quantities
\ben\lb{bd1}
r_*=\frac{ra_\infty^2}{GM}, \qquad s_*=\frac{V}{a_\infty},\qquad t_*=\frac\rho{\rho_\infty},
\een
which are related to the radial distance, flow velocity and mass density, respectively. Another dimensionless quantity which is useful in this analysis is the ratio of the flow velocity and the speed of sound $u_*=V/a$, which is the Mach number.

From the Newtonian mass density accretion rate we have in these dimensionless quantities
\ben\lb{bd2}
\dot M_\rho=4\pi \rho r^2V=4\pi\lambda\left(\frac{GM}{a_\infty^2}\right)^2\rho_\infty a_\infty,
\een
where $\lambda=r_*^2s_*t_*$ is a constant.

For the post-Newtonian approximation the Bondi dimensionless quantities  (\ref{bd1}) are written as
\ben\lb{bd3a}
r_*=\frac{ra_\infty^2}{GM}\left(1+\frac{GM}{2rc^2}\right),
\qquad
s_*=\frac{v_r}{a_\infty}\left(1+\frac{v_r^2}{2c^2}\right),\qquad t_*=\frac\rho{\rho_\infty}.
\een
  Solving (\ref{bd3a}) for $r$ and $v_r$ and considering terms up to $1/c^2$ we obtain
\ben\lb{bd3b}
r=\frac{GM}{a_\infty^2}r_*\left(1-\frac{\beta^2}{2r_*}\right),
\qquad
v_r={a_\infty}s_*\left(1-\frac{\beta^2}2s_*^2\right),\qquad
\een
where  $\beta={a_\infty}/{c}$ denotes a relativistic parameter which is the ratio of the value of the sound speed far from the massive object and the light speed.

With respect to the new variables (\ref{bd3a}),  the mass density accretion rate (\ref{s4a}) becomes
\ben\lb{bd3c}
\dot M_{\rho_*}=4\pi\lambda\left(\frac{GM}{a_\infty^2}\right)^2\rho_\infty a_\infty,
\een
by taking into account the Newtonian potential $\phi=-GM/r$. Here we have also that $\lambda=r_*^2s_*t_*$.

The dependence of the proper velocity as a function of the radial velocity is obtained from the Bernoulli equation (\ref{s7}) written in terms of the dimensionless quantities $(r_*,u_*)$. We begin by writing  the dimensionless parameters $s_*$ and $t_*$ as functions of  $(r_*,u_*)$
\ben\lb{bd4a}
s_*=u_*^{\frac2{\gamma+1}}\left(\frac{\lambda}{r_*^2}\right)^{\frac{\gamma-1}{\gamma+1}}\left[1+\frac{\beta^2}{\gamma+1}u_*^{\frac4{\gamma+1}}\left(\frac{\lambda}{r_*^2}\right)^{\frac{2(\gamma-1)}{\gamma+1}}\right],
\\\lb{bd4b}
t_*=\left(\frac\lambda{u_*r_*^2}\right)^{\frac2{\gamma+1}}\left[1-\frac{\beta^2}{\gamma+1}u_*^{\frac4{\gamma+1}}\left(\frac{\lambda}{r_*^2}\right)^{\frac{2(\gamma-1)}{\gamma+1}}\right],
\een
where we have taken into account  (\ref{bd3a})$_2$, $u_*=v_r/a$, $\lambda=r_*^2s_*t_*$ and $t_*=\rho/\rho_\infty=(a/a_\infty)^\frac2{\gamma-1}$.

Next we rewrite $v_r$ and $a$ in terms of $(r_*,u_*)$ from (\ref{bd3b})$_2$ and $a/a_\infty=t_*^\frac{\gamma-1}2$, yielding
\ben\lb{bd4c}
&&v_r=a_\infty u_*^{\frac2{\gamma+1}}\left(\frac{\lambda}{r_*^2}\right)^{\frac{\gamma-1}{\gamma+1}}\left[1-\frac{\beta^2(\gamma-1)}{2(\gamma+1)}u_*^{\frac4{\gamma+1}}\left(\frac{\lambda}{r_*^2}\right)^{\frac{2(\gamma-1)}{\gamma+1}}\right],
\\\lb{bd4d}
&&a=a_\infty \left(\frac{\lambda}{u_*r_*^2}\right)^{\frac{\gamma-1}{\gamma+1}}
\left[1-\frac{\beta^2(\gamma-1)}{2(\gamma+1)}u_*^{\frac4{\gamma+1}}\left(\frac{\lambda}{r_*^2}\right)^{\frac{2(\gamma-1)}{\gamma+1}}\right].
\een
The last step is to rewrite the gravitational potential $\phi$ and $\psi$ as functions of $r_*$
\ben\lb{bd4f}
\phi=-\frac{GM}r=-\frac{a_\infty^2}{r_*}\left(1+\frac{\beta^2}{2r_*}\right),
\qquad
\frac{\psi}{c^2}=-\frac{\phi^2}{4c^2}=-\frac{a_\infty^2}{4r_*^2}\beta^2.
\een
Note that in (\ref{bd4a}) -- (\ref{bd4f}) we have considered only terms up to the order $1/c^2$.

The final equation which gives the dependence of the Mach number $u_*$ with the dimensionless radial distance $r_*$ is obtained from Bernoulli equation (\ref{s7}) together with (\ref{bd4c}) -- (\ref{bd4f}) resulting
\ben\no
&&\frac{u_*^{\frac4{\gamma+1}}}2\left(\frac{\lambda}{r_*^2}\right)^{\frac{2(\gamma-1)}{\gamma+1}}\Bigg[1+\frac{\beta^2}{2(\gamma+1)}u_*^{\frac4{\gamma+1}}\Bigg(\frac{\lambda}{r_*^2}\Bigg)^{\frac{2(\gamma-1)}{\gamma+1}}\Bigg]-\frac1{r_*}\left(1+\frac{3\beta^2}{4r_*}\right)
\\\lb{bd5}
&&\qquad+\frac1{\gamma-1}\left(\frac{\lambda}{u_*r_*^2}\right)^{\frac{2(\gamma-1)}{\gamma+1}}\left[1-\left(1+\frac{(\gamma+1)}{2(\gamma-1)^2u_*^2}\right)\frac{\beta^2(\gamma-1)}{\gamma+1}u_*^{\frac4{\gamma+1}}\left(\frac{\lambda}{r_*^2}\right)^{\frac{2(\gamma-1)}{\gamma+1}}\right]
= \frac1{\gamma-1}\left(1-\frac{\beta^2}{2(\gamma-1)}\right).
\een

In the Newtonian limiting case we get -- by neglecting the $\beta^2$ terms -- eq. \emph{(14)} of Bondi \cite{4}, namely
\begin{equation}\lb{bd6}
\frac{u_*^{\frac4{\gamma+1}}}2\left(\frac{\lambda}{r_*^2}\right)^{\frac{2(\gamma-1)}{\gamma+1}}
+\frac1{\gamma-1}\left(\frac{\lambda}{u_*r_*^2}\right)^{\frac{2(\gamma-1)}{\gamma+1}}
=\frac1{r_*}+\frac1{\gamma-1}.
\end{equation}

In the next section we shall analyze the  relativistic spherically symmetrical accretion. This analysis will be based on the work of  Michel \cite{5} and on  the book by Shapiro and Teukolsky \cite{ST} .

\section{ Relativistic Accretion}\lb{sec5}

\subsection{Relativistic Bernoulli Equation}

We begin by writing  the  line element in spherical coordinates in the  Schwarzschild metric  $(r,\theta,\varphi)$
\be
ds^2=\left(1-\frac{2GM}{rc^2}\right)\left(dx^0\right)^2-\frac1{\left(1-\frac{2GM}{rc^2}\right)}\left(dr\right)^2-r^2\left[\left(d\theta\right)^2+\sin^2\theta\left(d\varphi\right)^2\right],
\ee{ssd1}
where $r_S=2GM/rc^2$ is the Schwarzschild radius  which defines the event horizon of a Schwarzschild black hole.

 The  perfect fluid  is  characterized by particle four-flow $N^\mu=nU^\mu$ and energy-momentum tensor (\ref{pf}) and the balance equations for the particle four-flow and energy-momentum tensor are given by
\ben\lb{ssd3a}
{N^\mu}_{;\mu}=\frac1{\sqrt{-g}}\frac{\partial\sqrt{-g}N^\mu}{\partial x^\mu}=0,
\\\lb{ssd3b}
{{T_{\mu}}^\nu}_{\,;\nu}=\frac1{\sqrt{-g}}\frac{\partial\sqrt{-g}\,{T_{\mu}}^\nu}{\partial x^\nu}-\frac12T^{\nu\sigma}\frac{\partial g_{\nu\sigma}}{\partial x^\mu}=0.
\een
In the  analysis of the spherically symmetrical accretion the non-vanishing components of the four-velocity are
\ben\lb{ssd4a}
\left(U^\mu\right)=\left(U^0=\frac{dx^0}{d\tau},U^r=\frac{dr}{d\tau},0,0\right).
\een
From  the constraint $g_{\mu\nu}U^\mu U^\nu=c^2$ the component $U^0$ is connected with $U^r$ by
\ben\lb{ssd4b}
\frac{U^0}c=\frac{\sqrt{1-\frac{2GM}{rc^2}+\left(\frac{U^r}c\right)^2}}{1-\frac{2GM}{rc^2}}.
\een

The integration of the balance equation for the particle four-flow (\ref{ssd3a}) and the time component of the energy-momentum tensor (\ref{ssd3b}) lead to
\ben\lb{ssd5a}
\sqrt{-g}nU^r=\hbox{constant},\qquad
\sqrt{-g}\left(p+\epsilon\right)\frac{U^r}{c}\frac{U_0}{c}=\hbox{constant}.
\een
Combining the above equations  the following relationship holds
\ben\lb{ssd7b}
\left(\frac{p+\epsilon}{\rho}\right)^2\left[1-\frac{2GM}{rc^2}+\left(\frac{U^r}c\right)^2\right]=\hbox{constant}.
\een

Let us introduce  the  sound speed $a$, which for a relativistic fluid is defined by
\ben\lb{ssd8a}
\frac{a^2}{c^2}=\frac\rho{p+\epsilon}\left(\frac{\partial p}{\partial\rho}\right).
\een
We recall that the polytropic equation of state and the energy density equation are given by
\ben
p=\kappa\rho^\gamma,\qquad \epsilon=\rho c^2+\frac{\kappa\rho^\gamma}{\gamma-1},
\een
so that we can write  from the above equations that
\ben
\frac{p+\epsilon}\rho=\frac{\kappa\gamma\rho^{\gamma-1}}{\gamma-1}+c^2=\frac{\kappa\gamma\rho^{\gamma-1}}{a^2/c^2},
\een
which implies the following  relationships
\ben\lb{ssd9b}
\kappa\gamma\rho^{\gamma-1}=\frac{(\gamma-1)a^2}{\gamma-1-a^2/c^2},\qquad\frac{p+\epsilon}\rho=\frac{c^2}{1-a^2/(\gamma-1)c^2}.
\een

The relativistic Bernoulli equation follows from  (\ref{ssd7b}) and (\ref{ssd9b}), yielding
\ben\lb{ssd10a}
\left(1-\frac{a^2}{c^2(\gamma-1)}\right)^2=\left(1-\frac{a_\infty^2}{c^2(\gamma-1)}\right)^2\left[1-\frac{2GM}{rc^2}+\left(\frac{U^r}c\right)^2\right].
\een
Here  it was supposed that far from the massive body
${2GM}/{rc^2}$ and $U^r$ vanish while the sound speed becomes $a_\infty$.

The determination of  the critical points are obtained from the differentiation of  (\ref{ssd5a}) and elimination of $d\rho$, yielding
\ben\no
&&\frac{dU^r}{U^r}\left[\frac{a^2}{c^2}-\frac{\left(U^r/c\right)^2}{1-{2GM}/{rc^2}+\left({U^r}/{c}\right)^2}\right]
\\\lb{ssd11}
&&\qquad+\frac{dr}r\left[2\frac{a^2}{c^2}-\frac{GM/rc^2}{1-{2GM}/{rc^2}+\left({U^r}/{c}\right)^2}\right]=0.
\een
The expressions for the critical gas flow velocity and sound speed are determined when both expressions in the parenthesis in (\ref{ssd11}) vanish resulting
\be
({U^r_c})^2={\frac{GM}{2r_c}},\qquad a_c^2={\frac{({U^r_c})^2}{1-3(U^r_c/c)^2}},\qquad ({U^r_c})^2={\frac{a_c^2}{1+3(a_c/c)^2}}.
\ee{ssd12}
The above equations correspond to the equations \emph{(8)} -- \emph{(14)} of the work of Michel \cite{5}.

From now one we shall restrict the analysis to the  weak field limit of the relativistic case, since we are interested in comparing it  with the post-Newtonian approximation developed in the previous section. We begin by writing the Bernoulli equation  (\ref{ssd10a}) at the critical point  thanks to (\ref{ssd12}) as
\ben
\left(1+3\frac{a_c^2}{c^2}\right)\left(1-\frac{a_c^2}{c^2(\gamma-1)}\right)^2=\left(1-\frac{a_\infty^2}{c^2(\gamma-1)}\right)^2,
\een
which is a third order algebraic equation for the determination of the critical sound speed $a_c^2$. This equation was solved in \cite{SHA} but here we are interested in its weak field approximation which reads
\ben
a_c^2=\frac{2a_\infty^2}{5-3\gamma}\left[1-\frac{3(3\gamma+1)}{2(5-3\gamma)(\gamma-1)}\frac{a_\infty^2}{c^2}\right].
\een

From the knowledge of the critical sound speed the critical values for the flow velocity, mass density  and radial distance  read
\ben
({U^r_c})^2=\frac{2a_\infty^2}{5-3\gamma}\left[1+\frac{3(7-11\gamma)}{4(5-3\gamma)(\gamma-1)}\frac{a_\infty^2}{c^2}\right],
\\
\frac{\rho_c}{\rho_\infty}=\left(\frac{2}{5-3\gamma}\right)^\frac1{\gamma-1}\left[1-\frac{3(3\gamma+1)}{2(5-3\gamma)(\gamma-1)^2}\frac{a_\infty^2}{c^2}\right],
\\
r_c=\frac{(5-3\gamma)}4\frac{GM}{a_\infty^2}\left[1-\frac{3(7-11\gamma)}{4(\gamma-1)(5-3\gamma)}\frac{a_\infty^2}{c^2}\right].
\een
Furthermore, from the mass accretion rate $\dot M=4\pi r_c^2\rho_c V_c$ it follows the critical accretion eigenvalue
\ben
\lambda_c=\left(5-3\gamma\right)^\frac{3\gamma-5}{2\gamma-2}2^\frac{9-7\gamma}{2\gamma-2}\left[1+\frac{3(17-66\gamma+33\gamma^2)}{8(\gamma-1)^2(5-3\gamma)}
\frac{a_\infty^2}{c^2}\right].
\een
Note that the above expressions differ from those obtained in the post-Newtonian approximation.

 From the relativistic Bernoulli equation  one may obtain its weak field limit  by considering terms up to the $1/c^2$ order in (\ref{ssd10a}), yielding
\ben\no
&&\frac{(U^r)^2}{2}\left[1-\left(\frac{U^r}{c}\right)^2-\frac{4\phi}{c^2}\right]+\frac{a^2}{(\gamma-1)}\left[1-\frac{a^2}{2c^2(\gamma-1)}-\frac{2\phi}{c^2}-\left(\frac{U^r}{c}\right)^2\right]
\\\lb{ssd10b}
&&\qquad+\phi\left(1-\frac{2\phi}{c^2}\right)=\frac{a_\infty^2}{(\gamma-1)}\left(1-\frac{a_\infty^2}{2c^2(\gamma-1)}\right),
\een
where we have introduced the Newtonian potential $\phi=-GM/r$. Without the $1/c^2$ -- terms (\ref{ssd10b})  reduces to the non-relativistic Bernoulli equation, however this expression  differs from the post-Newtonian Bernoulli equation (\ref{s7}).

Let express the weak filed approximation of the Bernoulli equation in terms of the proper velocity of the flow $v_r$ defined by (\ref{nn1}).
The relationship between the components $U^r$ and $U^0$ follows from $U_\mu U^\mu=c^2$, yielding
\ben\lb{nn2}
\left(\frac{U^0}{c}\right)^2=\frac{1+\left(1-\frac{2\phi}{c^2}\right)\left(\frac{U^r}{c}\right)^2}{1+\frac{2\phi}{c^2}},
\een
and the proper velocity  (\ref{nn1}) becomes
\ben\lb{nn3}
v_r=\frac{U^r}{\sqrt{1+2\frac{\phi}{c^2}}\sqrt{1+\left(1-\frac{2\phi}{c^2}\right)\left(\frac{U^r}{c}\right)^2}}.
\een
 By retaining terms up to $1/c^2$ the expression of the radial four-velocity component  in terms of the proper velocity reads
\ben\lb{nn4}
U^r=v_r\left[1+\frac{\phi}{c^2}+\frac{v_r^2}{2c^2}\right].
\een

If we insert  (\ref{nn4}) into (\ref{ssd10b}) and consider terms up to $1/c^2$ order we find the weak field Bernoulli equation given in terms of the proper velocity, namely
\ben\no
&&\frac{v_r^2}{2}\left[1-\frac{2\phi}{c^2}\right]+\frac{a^2}{(\gamma-1)}\left[1-\frac{a^2}{2c^2(\gamma-1)}\underline{-\frac{2\phi}{c^2}-\frac{v_r^2}{c^2}}\right]+\phi\left(1-\frac{2\phi}{c^2}\right)-\frac{a_\infty^2}{(\gamma-1)}\left(1-\frac{a_\infty^2}{2c^2(\gamma-1)}\right)
\\\lb{nn5}
&&\qquad=\frac{v_r^2}{2}\left[1-\frac{2\phi}{c^2}\right]+\frac{a^2}{(\gamma-1)}\left[1+\frac{3a^2-4a_\infty^2}{2c^2(\gamma-1)}\right]+\phi\left(1-\frac{2\phi}{c^2}\right)-\frac{a_\infty^2}{(\gamma-1)}\left(1-\frac{a_\infty^2}{2c^2(\gamma-1)}\right)=0.
\een
 For the underlined term above we have used the Newtonian Bernoulli equation (\ref{s5b}), since it is of $1/c^2$ order.

\subsection{Mach Number as Function of the Radial Distance}

The mass density accretion rate for the weak field is obtained from (\ref{ssd5a}) which in terms of the proper velocity reads
\ben\lb{po1a}
\dot M=4\pi\rho r^2 U^r=4\pi\rho  r^2 v_r\left[1-\frac{GM}{rc^2}+\frac{v_r^2}{2c^2}\right].
\een

Following the same methodology of the previous section we introduce the dimensionless quantities
\ben\lb{po1b}
r_*=\frac{r a_\infty^2}{GM}\left(1-\frac{GM}{2rc^2}\right), \qquad s_*=\frac{v_r}{a_\infty}\left(1+\frac{v_r^2}{2c^2}\right), \qquad t_*=\frac\rho{\rho_\infty},
\een
so that the mass density accretion rate becomes
\ben\lb{po1c}
\dot M=4\pi\lambda\left(\frac{GM}{a_\infty^2}\right)^2\rho_\infty a_\infty, \qquad\hbox{where}\qquad \lambda=r_*^2s_*t_*.
\een
From (\ref{po1b}) we can write
\ben\lb{po1d}
r=\frac{GMr_*}{a_\infty^2}\left(1+\frac{\beta^2}{2r_*}\right),\qquad v_r=a_\infty s_*\left(1-\frac{\beta^2s_*}2\right).
\een

Due to the fact that the expression for $s_*$ above  is the same as the one in the post-Newtonian approximation (\ref{bd3a}) we can use the (\ref{bd4c}) and (\ref{bd4d}) for the proper velocity and sound speed as a function of the Mach number $u_*$ and dimensionless radial distance $r_*$, respectively. For the gravitational potential we have
\ben\lb{po2a}
\phi=-\frac{a_\infty^2}{r_*}\left(1-\frac{\beta^2}{2r_*}\right).
\een

The expressions of the dependence of the Mach number $u_*$ as function of the dimensionless radial distance $r_*$ is obtained from the weak field Bernoulli equation (\ref{nn5})  together with (\ref{bd4c}) , (\ref{bd4d}) and (\ref{po2a}), resulting
\ben\no
&&\frac{u_*^{\frac4{\gamma+1}}}2\left(\frac{\lambda}{r_*^2}\right)^{\frac{2(\gamma-1)}{\gamma+1}}\Bigg[1-\frac{\beta^2(\gamma-1)}{\gamma+1}u_*^{\frac4{\gamma+1}}\Bigg(\frac{\lambda}{r_*^2}\Bigg)^{\frac{2(\gamma-1)}{\gamma+1}}+\frac{2\beta^2}{r_*}\Bigg]-\frac1{r_*}\left(1+\frac{3\beta^2}{2r_*}\right)+\frac1{\gamma-1}\left(\frac{\lambda}{u_*\,r_*^2}\right)^{\frac{2(\gamma-1)}{\gamma+1}}\Bigg[1-\frac{2\beta^2}{\gamma-1}
\\\lb{ber2}
&&\qquad-\left(1-\frac{3(\gamma+1)}{2(\gamma-1)^2u_*^2}\right)\frac{\beta^2(\gamma-1)}{(\gamma+1)}u_*^{\frac4{\gamma+1}}\left(\frac{\lambda}{r_*^2}\right)^{\frac{2(\gamma-1)}{\gamma+1}}
\Bigg]
= \frac1{\gamma-1}\left(1-\frac{\beta^2}{2(\gamma-1)}\right).
\een

For the relativistic Bernoulli equation (\ref{ssd10a}) we can use the Bondi dimensionless quantities (\ref{bd1}) and write it as
\ben\lb{ber3a}
\left[1-\frac{\beta^2}{\gamma-1}\left(\frac{\lambda}{U_*r_*^2}\right)^\frac{2(\gamma-1)}{\gamma+1}\right]^2=\left(1-\frac{\beta^2}{\gamma-1}\right)^2\left[1-\frac{2\beta^2}{r_*}+\beta^2U_*^2\left(\frac{\lambda}{U_*r_*^2}\right)^\frac{2(\gamma-1)}{\gamma+1}\right].
\een
Here $U_*=U^r/a$ is the Mach number with respect to the radial component of the four-velocity. The Mach number written in terms of the proper velocity follows from (\ref{nn3}) and reads
\ben\lb{ber3b}
u_*=\frac{v_r}a=\frac{U_*}{\sqrt{1-2\frac{\beta^2}r}\sqrt{1+\left(1+2\frac{\beta^2}r\right)\beta^2U_*^2\left(\frac{\lambda}{U_*r_*^2}\right)^\frac{2(\gamma-1)}{\gamma+1}}}.
\een

\section{Analysis of the solutions}\lb{sec6}

In this section we shall compare the solutions for the Mach number $u_*=v_r/a$ as function of the dimensionless radial distance $r_*$ which follow from the different approximations of the Bernoulli equation.

 In the tables and figures below the Newtonian solution of  (\ref{bd6}) is denoted by (N), the post-Newtonian solution of  (\ref{bd5})  by (PN) and  the weak field approximation solution of   (\ref{ber2}) by (WF).
For the relativistic accretion -- denoted by  (R) -- the Bernoulli equation (\ref{ber3a}) was solved for the Mach number with respect to the radial four-velocity and from (\ref{ber3b}) the Mach number for the proper velocity was obtained.

 In the determination of the Mach number $u_*$ as a function of the dimensionless radial distance $r_*$ it was  considered that the ratio of the sound velocity far from the massive body and the  light speed  is equal to $\beta=a_\infty/c=10^{-2}$, which is of relativistic order.

\begin{figure}[ht]
\centerline{\includegraphics[width=10cm]{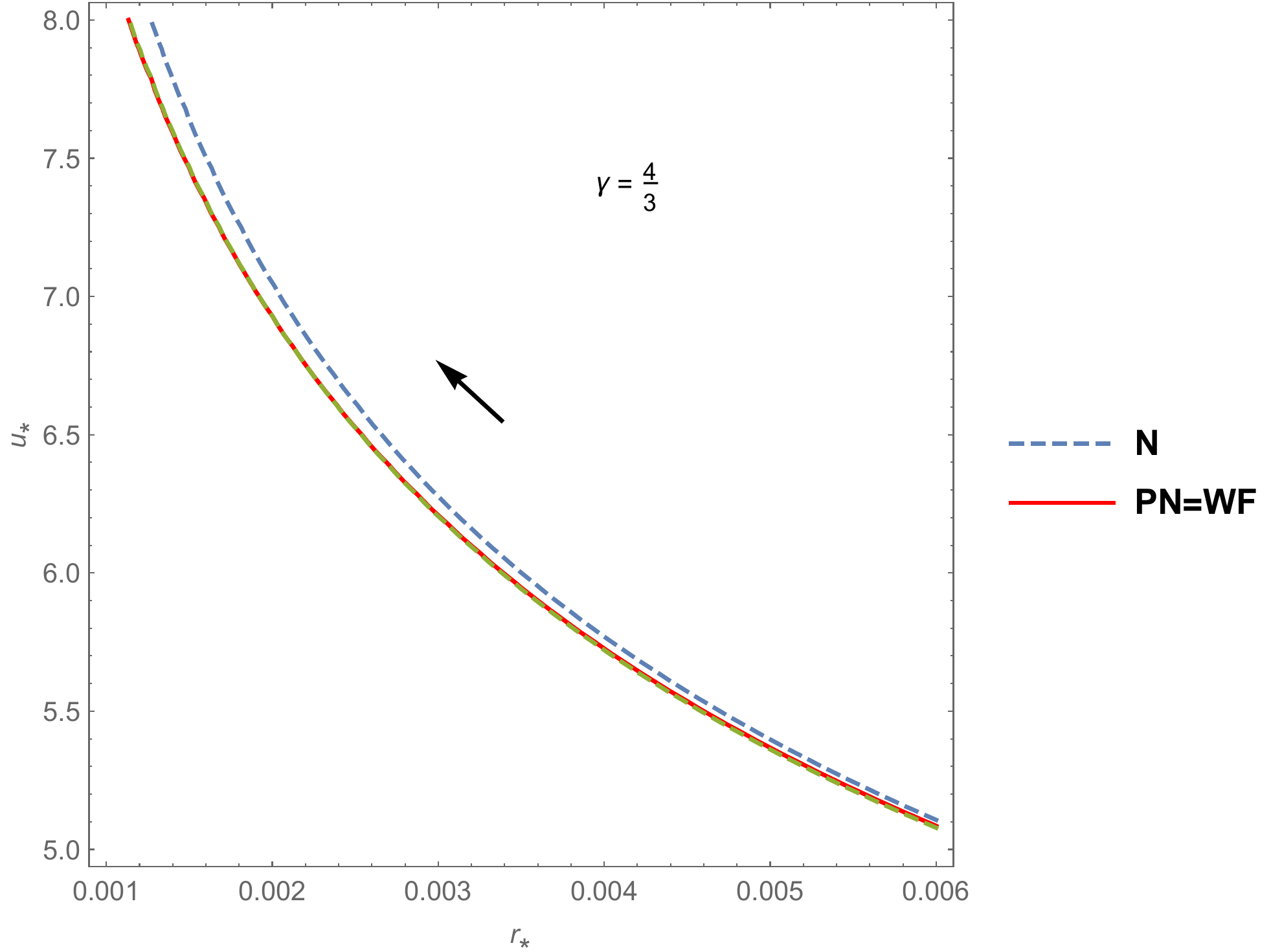}}
\caption{Contour plot showing the Mach number $u_*$ as function of the dimensionless radial distance $r_*$ for a ultra-relativistic Fermi gas $\gamma=4/3$. The  Newtonian solution is represented by a dashed  line while the  post-Newtonian and weak field solutions by a straight line.}\label{ff.1}
\end{figure}

In Table \ref{t1.1} the values for the Mach number $u_*$ as function of the dimensionless radial distance $r_*$ are given in the range $5\times 10^{-4}\leq r_*\leq2.5\times 10^{-2}$   for a ultra-relativistic Fermi gas  where $\gamma=4/3$.
In the Newtonian approximation the critical radius is $r_*=0.25$ where the critical Mach number assumes the value $u_*=1$. We infer from this table that by decreasing the  dimensionless radial distances $r_*$ from the massive body the Mach number increases. Furthermore, the values of the Mach number for the relativistic case  are bigger than the Newtonian ones. The Mach number values for  the post-Newtonian and weak field approximations are practically the same  and are smaller than those  for the Newtonian case.  The difference between the Newtonian, post-Newtonian and weak field solutions  becomes very small by increasing the dimensionless radial distance and the solutions practically coincide at $r_*=10^{-3}$. In Figure \ref{ff.1} it is plotted the  contour plot  for the Bernoulli equations: Newtonian  (\ref{bd6}),  post-Newtonian  (\ref{bd5})  and weak field (\ref{ber2}). The Newtonian solution is represented by a dashed line and the post-Newtonian and weak field approximations by the same straight line, since they practically coincide. It is shown that the difference between the Newtonian, post-Newtonian and weak field are very small and coincide by increasing the dimensionless radial distance.

\begin{table}[ht]
 \begin{tabular}{|c|c|c|c|c|c|}\hline
$r_*$&$u_*$ (N)&$u_*$ (PN)&$u_*$ (WF)&$u_*$ (R)\\\hline
$5\times10^{-4}$&10.29&9.66&9.66&20.26\\
$10^{-3}$&8.53&8.24&8.26&11.26\\
$5\times10^{-3}$&5.40&5.37&5.37&6.07\\
$10^{-2}$&4.35&4.34&4.34&4.97\\
$5\times10^{-2}$&2.40&2.41&2.41&3.31\\
$2.5\times10^{-2}$&1.00&1.07&1.02&2.09\\
        \hline
    \end{tabular}
     \caption{Mach number $u_*=v_r/a$ as function of the dimensionless radial distance $r_*$ for a ultra-relativistic Fermi gas $\gamma=4/3$.}
        \label{t1.1}
       \end{table}

The Mach number $u_*$ as function of the dimensionless radial distance $r_*$  for a diatomic gas where $\gamma=7/5$ is displayed in Table \ref{t1.2}  in the range $5\times 10^{-4}\leq r_*\leq2\times 10^{-2}$. The same conclusions as in the former case can be drawn, i.e., in comparison with the Newtonian solutions the dependence of Mach number with respect to the dimensionless radial distance for the relativistic case  is bigger, the post-Newtonian and the weak field solutions are smaller and both  have practically the same values. For the Newtonian case the critical radius is $r_*=0.2$ where the Mach number attains the value $u_*=1$.  In Figure 2 the contour plots of the Newtonian (dotted line), the post-Newtonian (straight line)  and the weak field (dotted line) solutions are displayed showing that the values of the Mach number for the Newtonian solution is bigger that those for the post-Newtonian and weak field solutions and that the difference between them  becomes very small by increasing the dimensionless radial distance.

\begin{table}[ht]
 \begin{tabular}{|c|c|c|c|c|c|}\hline
$r_*$&$u_*$ (N)&$u_*$ (PN)&$u_*$ (WF)&$u_*$ (R)\\\hline
$5\times10^{-4}$&7.21&6.83&6.81&14.22\\
$10^{-3}$&6.16&5.99&5.98&8.22\\
$5\times10^{-3}$&4.16&4.14&4.13&4.77\\
$10^{-2}$&3.43&3.43&3.43&4.02\\
$5\times10^{-2}$&2.00&2.00&2.00&2.86\\
$2\times10^{-2}$&1.00&1.02&1.06&2.21\\
        \hline
    \end{tabular}
     \caption{Mach number $u_*=V/a$ as function of the dimensionless radial distance $r_*$ for a diatomic gas $\gamma=7/5$.}
        \label{t1.2}
       \end{table}

\begin{figure}[ht]
\centerline{\includegraphics[width=10cm]{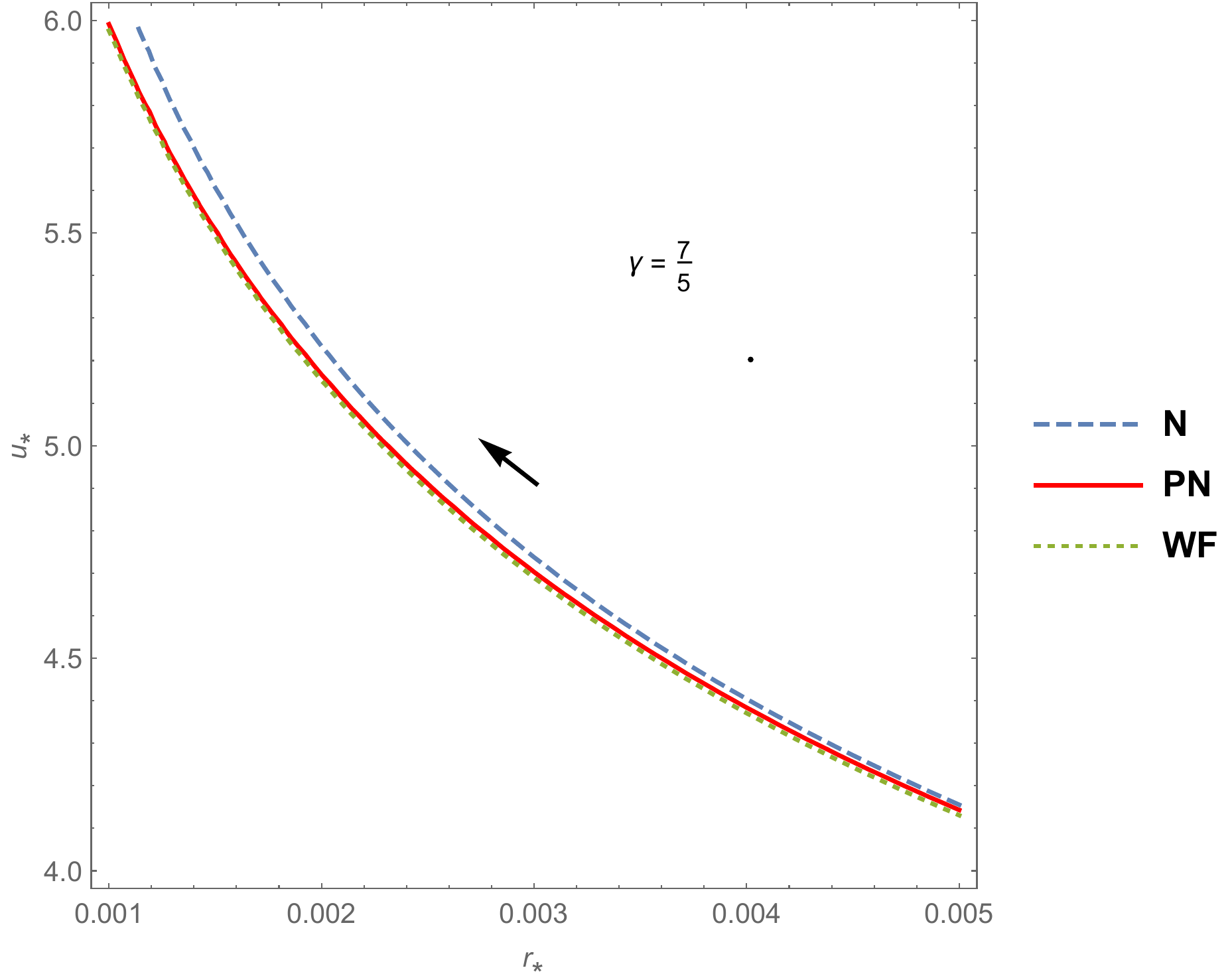}}
\caption{Contour plot showing the Mach number $u_*$ as function of the dimensionless radial distance $r_*$ for a diatomic gas $\gamma=7/5$. The  Newtonian solution is represented by a dashed  line, the  post-Newtonian by straight line and weak field by a dotted  line. }\lb{ff.2}
\end{figure}

\begin{figure}[ht]
\centerline{\includegraphics[width=10cm]{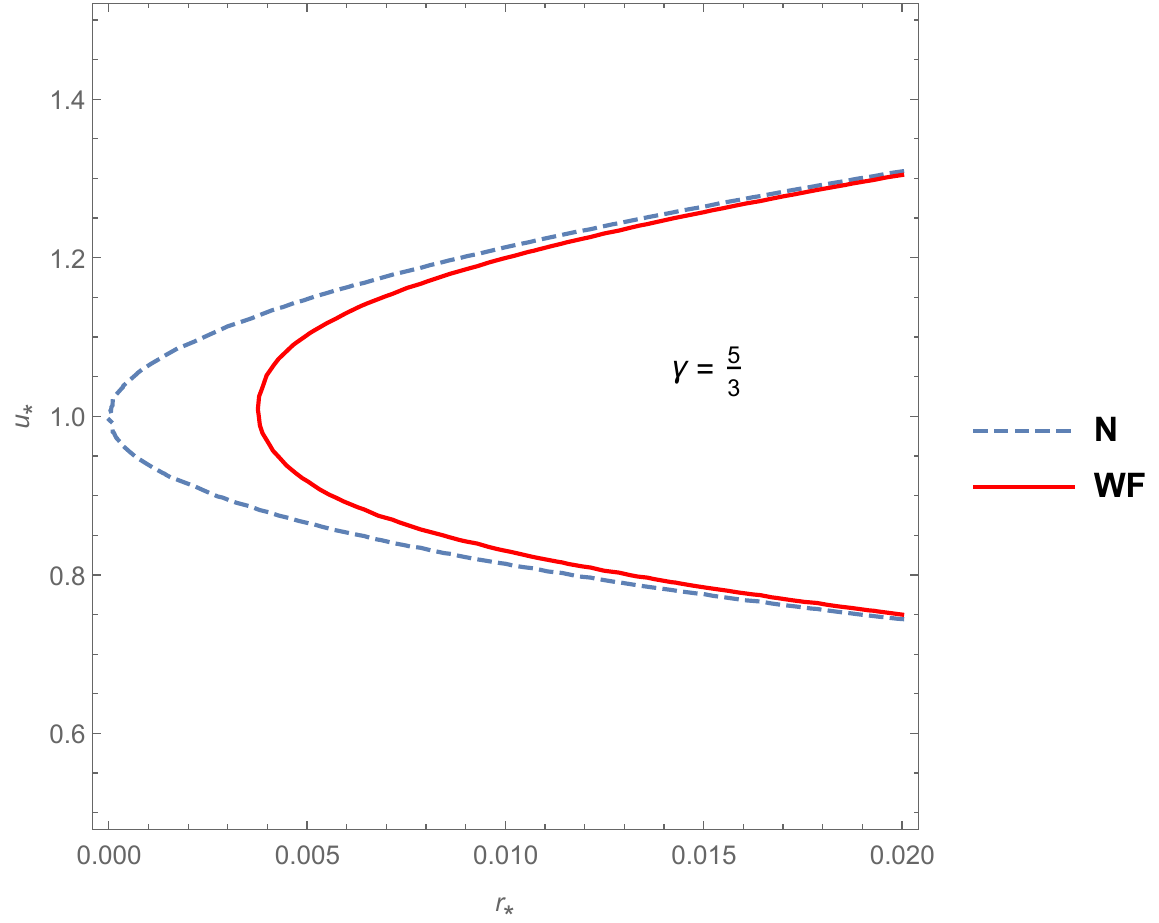}}
\caption{Contour plot showing the Mach number $u_*$ as function of the dimensionless radial distance $r_*$ for a non-relativistic Fermi gas $\gamma=5/3$. The  Newtonian solution is represented by a dashed  line while the  weak field solution by a straight line. }\label{ff.3}
\end{figure}

For a non-relativistic Fermi gas or a monatomic gas  $\gamma=5/3$ and in this case  the contour plots of the Newtonian and weak field solutions are shown in Figure \ref{ff.3}. The critical dimensionless radial distance for the Newtonian case is $r_*=0$ where the the Mach number becomes equal to $u_*=1$.  We note that $r_*=0$   is a turning point for the Newtonian solution where a  transition occurs from an accretion flow  to a wind flow.  The weak field solution is smaller than the Newtonian ones and the  turning point   is about  $r_*\approx 4\times 10^{-3}$.

Here it is important to comment the behaviors of the post-Newtonian and weak field solutions found in the above analysis when compared with the Newtonian and relativistic solutions. It was expected that  the post-Newtonian and weak field solutions should be more close to the relativistic one and not smaller than the Newtonian solution. By inspecting the Newtonian     (\ref{bd6}), the post-Newtonian  (\ref{bd5})  and  the weak field    (\ref{ber2}) equations we infer that the two latter equations have corrections from the Newtonian one and their solutions should furnish different results for the dependence of the Mach number as function of the dimensionless radial distance. But why the values of the Mach number for the post-Newtonian and weak field  are smaller than in the Newtonian case? The only clue is to look at the expression for the proper velocity (\ref{bd4c}) for the post-Newtonian and weak field  which can be written as
\ben
v_r=a_\infty u_*^{\frac2{\gamma+1}}\left(\frac{\lambda}{r_*^2}\right)^{\frac{\gamma-1}{\gamma+1}}\left[1-\frac{\beta^2(\gamma-1)}{2(\gamma+1)}u_*^{\frac4{\gamma+1}}\left(\frac{\lambda}{r_*^2}\right)^{\frac{2(\gamma-1)}{\gamma+1}}\right]=v_r^N\left[1-\frac{\beta^2(\gamma-1)}{2(\gamma+1)}u_*^{\frac4{\gamma+1}}\left(\frac{\lambda}{r_*^2}\right)^{\frac{2(\gamma-1)}{\gamma+1}}\right],
\een
where $v_r^N$ is the Newtonian expression for the proper velocity. One infers from the above equation that the proper velocities for the post-Newtonian and weak field should be  smaller than the one for the Newtonian case, which could explain the difference in the behavior of the solutions.

\section{Summary}\lb{sec7}

In this work we have analyzed the influence of the first post-Newtonian approximation in the spherical symmetrical accretion of an infinity  gas cloud  characterized by a polytropic equation of state into a massive object. The starting point was the steady state post-Newtonian hydrodynamics equations for mass, mass-energy and momentum densities. The  integration of the system of equations in spherical coordinates --  where the fields depend only on the radial coordinate -- lead to the determination of the mass accretion rate and the Bernoulli equation in the post-Newtonian approximation. From the system of differential equations for mass density, flow velocity and post-Newtonian potentials the critical point was identified. The critical point prevent singularities in the flow solution and guarantees a smooth monotonic increase of the flow velocity along the trajectory of the particle so that through the critical point  a continuous inflow and outflow velocity happen. The critical point in the accretion Newtonian theory corresponds to the transonic point where the flow velocity matches the sound speed. In the post-Newtonian approximation the critical flow velocity is connected with the sound speed but their expression are not the same. From the post-Newtonian Bernoulli equation an equation for the Mach number was obtained as a function of a dimensionless radial coordinate. Similar expressions were derived for the relativistic Bernoulli equation and its weak field approximation based on the work by Michel \cite{5}. For the solution of the post-Newtonian equation  it was considered that the ratio of the sound velocity far the massive body and the speed of light was of order $a_\infty/c=10^{-2}$ which is of relativistic order. The results obtained were: (i)  the Mach number for the Newtonian, post-Newtonian and weak field accretions have practically the same values for radial distances of order of the critical radial distance; (ii) by decreasing the radial distance the Mach number for the Newtonian accretion is bigger than the one for the  post-Newtonian and weak field  accretions; (iii) the effect of the correction terms in post-Newtonian and weak field Bernoulli equations are more perceptive for the lowest values of the radial distance; (iv) practically there is no difference between the Newtonian, post-Newtonian and weak field Mach numbers when the ratio $a_\infty/c\ll10^{-2}$; (v) the solutions for $a_\infty/c>10^{-2}$ does not lead to   a continuous inflow and outflow velocities at the critical point; (vi) from the comparison of the solutions with those that follow from the relativistic Bernoulli equation shows that the Mach number of the former is bigger than the Newtonian, post-Newtonian and weak field Mach numbers.

\section*{Acknowledgments}

G. M. K. has been supported by  CNPq (Conselho Nacional de Desenvolvimento Cient\'ifico e Tecnol\'ogico), Brazil and L. C. M.  by CAPES (Coordena\c c\~ao de Aperfei\c coamento de Pessoal de N\'ivel Superior), Brazil. We thank the referee for suggestions and comments.

\end{document}